\documentclass[11pt]{article}
\usepackage{putex}
\usepackage[colorinlistoftodos]{todonotes}
\usepackage{tcolorbox}
\usepackage{amsfonts}
\usepackage{amssymb}
\usepackage{amsmath}
\usepackage[bookmarks = false, colorlinks = true, linkcolor = blue, citecolor = purple]{hyperref}
\usepackage{ulem}
\usepackage{tipa}
\usepackage{graphics,graphicx}
\usepackage{setspace}
\numberwithin{equation}{section}

\newcommand{\HH}{\mathcal{H}}

\newcommand{\AAA}{\mathcal{A}}

\newcommand{\Tr}{\mbox{Tr}}

\newcommand\be{\begin{equation}}
\newcommand\ba{\begin{eqnarray}}
\newcommand\ee{\end{equation}}
\newcommand\ea{\end{eqnarray}}
\newcommand\bone{{\bf 1}}
\definecolor{purple}{rgb}{0.7,0.0,0.5}
\definecolor{huh}{rgb}{0.0,0.6,0.8}
\definecolor{orange}{rgb}{1,0.5,0}
\definecolor{pink}{rgb}{1,0.4,0.4}
\definecolor{light-gray}{gray}{0.8}

\begin{document}

\title{Entanglement entropy in Jackiw-Teitelboim Gravity}
\authors{Jennifer Lin}
\institution{IAS}{{School of Natural Sciences, Institute for Advanced Study, Princeton, NJ, USA}}
\abstract{
I show that the black hole entropy associated to an $AdS_2$ wormhole is an entanglement edge term related to a natural measure on the gauge group in the $SL(2)$ gauge theory formulation of $1+1d$ Jackiw-Teitelboim gravity. I comment on what the entropy appears to be counting.} 

\maketitle
\newpage

\section{Introduction}\label{s1}

We would like to understand the universal origin of the black hole entropy, 
\be\label{bhe}
S_{BH} = \frac{A}{4G_N}\,.
\ee

A related and perhaps easier problem is to understand why the Ryu-Takayanagi formula in AdS/CFT \cite{Ryu:2006bv}, \cite{Ryu:2006ef},
\be\label{rt}
S_{EE}(A) = \frac{A_{min}(A)}{4G_N} + S_{alg, ginv}(\mathcal{E}_A)
\ee
for $A$ a subregion in a state of a holographic CFT with an Einstein gravity dual, $A_{min}(A)$ the area of the minimal-area homologous bulk surface, and $S_{alg,ginv}(\mathcal{E}_A$) the algebraic EE of gauge-invariant operators in the entanglement wedge of $A$ \cite{Faulkner:2013ana, Dong:2016eik, Harlow:2016vwg}, is true from the bulk point of view. 

In \cite{Lin:2017uzr} (see also \cite{Donnelly:2016auv}), I pointed out that the RT formula  resembles a formula for entanglement entropy of a spatial subregion in an emergent gauge theory \cite{Buividovich:2008gq, Donnelly:2011hn, Ghosh:2015iwa, Radicevic:2015sza}, which is
\begin{eqnarray}\label{lgtee}
S_{EE, UV}(A) &=& \sum_R p_R\left[-\log p_R + \log \dim R\right]+ \mbox{EE distillable with g-inv operations} + c \\ &=& \sum_R p_R \log \dim R +  S_{alg, ginv}(A) +  c\,,\label{lgtee2}
\end{eqnarray}
on the lattice, \footnote{This equation also makes sense in the continuum but it takes a bit more space to explain what the terms mean.}
where the index $R$ labels the distribution of boundary link representations in a state of the lattice gauge theory, $p_R$ is the normalized probability distribution over these superselection sectors, and the constant $c$ is independent of the IR state. 
The first and second terms of \eqref{lgtee} are sometimes called the Shannon and $\log \dim R$ edge terms. In particular, we identify
\be\label{conjecture}
\frac{A}{4G_N} \sim \log \dim R \mbox{ edge term}\,.
\ee
This would be quite interesting if true because in an ordinary gauge theory we know what the various components of \eqref{lgtee} are counting at the level of the operator algebra. 
\footnote{
To forestall a possible confusion, the $\log \dim R$ edge term is associated to the entanglement of static background charges as opposed to operators in the IR, as I will explain below. So the results reported here are consistent with the non-factorizability of the JT Hilbert space \cite{Harlow:2018tqv}.
}
We might be able to use this knowledge to help us understand what the black hole entropy is counting from the bulk point of view. But so far it was just an analogy, obtained by applying \eqref{lgtee} well outside its regime of validity.

In this note, I will show explicitly that a version of \eqref{lgtee}, \eqref{conjecture} is true in $1+1d$ Jackiw-Teitelboim (JT) gravity. 
JT gravity has a first order formulation as a topological BF gauge theory with gauge group $SL(2,R)$.
I show that in the gauge theory formulation, the entanglement across the midpoint of the interval in the Hartle-Hawking state at inverse temperature $\beta$, computed with the replica trick, can be put in the form
\be\label{result1}
S_{EE}(\Psi^{HH}_\beta)  = \int_k p_{k, \beta}[-\log p_{k, \beta} + \log k \sinh 2\pi k ]\,,
\ee
where $p_{k, \beta}$ is the normalized probability distribution over the representation basis of the BF Hilbert space in the Hartle-Hawking state; $``k \sinh 2\pi k"$ is a Plancherel measure for $SL(2,R)$, that naturally generalizes $``\dim R"$ for infinite-dimensional reps; and there is no distillable EE. In the classical limit, \eqref{result1} is dominated by the second term which agrees with the JT analog of \eqref{bhe} when evaluated on the peak of the Hartle-Hawking wavefunction,
\be\label{result2}
S_{EE}(\Psi^{HH}_\beta) = \left. \log k \sinh 2\pi k\right|_{k_{max}(\beta)} = \frac{\langle\phi_h\rangle}{4G_N}\,,
\ee
for $\langle\phi_h\rangle$ the expectation value of the dilaton at the bifurcate horizon.
This demonstrates \eqref{conjecture} in JT gravity.
By comparing to the situation in compact gauge theories, I suggest what the entropy might be counting.

In section \ref{s2} I review some relevant aspects of JT gravity. In section \ref{s3} I show the main results. I suggest an interpretation for what we are counting in section \ref{s4}, and conclude in section \ref{s5}. An appendix contains a brief introduction to the subject of entanglement entropy in gauge theories.

\section{Review of Jackiw-Teitelboim gravity}\label{s2}

2d Jackiw-Teitelboim gravity \cite{Jackiw:1984je}, \cite{Teitelboim:1983ux} is the simplest nontrivial theory of gravity. It also happens to be an emergent theory of gravity. Hence, it's a natural place to study the conjecture \eqref{conjecture}. 

JT gravity on a line interval is characterized by the action 
\be\label{jtaction}
S_{JT} = 
\frac{1}{16\pi G_N}\left[\int_M d^2x \sqrt{g} \phi (R +2) + 2 \int_{\partial M} dt\, \phi K
\right]\,.
\ee
The first term is the bulk action and the second is the GHY boundary term needed to make the variational problem well-defined. 
To simplify the notation below, I will sometimes use
\be
\alpha = \frac{1}{16\pi G_N}\,.
\ee

The equations of motion are  
\begin{eqnarray}
R+2 &=& 0\,, \label{eom1} \\
(\nabla_\mu \nabla_\nu - g_{\mu\nu})\phi &=& 0\,. \label{eom2}
\end{eqnarray}

In addition, we have to choose boundary conditions such that the boundary term vanishes. 
One natural choice is to fix Dirichlet bc's at a cutoff surface,
\be\label{jtbc}
\left. g\right|_{\partial M} = \frac{1}{\epsilon^2}, \qquad \left. \phi\right|_{\partial M} = \frac{\phi_b}{\epsilon}\,,
\ee
and then send $\epsilon \rightarrow 0$ with $\phi_b$ fixed and positive.

The solutions to \eqref{eom1} - \eqref{jtbc} were recently discussed in \cite{Harlow:2018tqv}. 
There is a one-parameter family of $AdS_2$ wormhole solutions labeled by the value of $\phi = \phi_h$ at the bifurcate horizon. When we quantize the theory, these correspond to a Hilbert space basis of Hamiltonian eigenstates.
Another natural one-parameter family of states are the Hartle-Hawking states labeled by $\beta$, the inverse temperature of the black hole as seen by an outside observer. 
These states were constructed in \cite{Harlow:2018tqv} in the gravity variables and I will construct them in the BF variables below.

\subsection{Emergence from the SYK model}

The SYK model (\cite{Sachdev:1992fk, Kitaev, Maldacena:2016hyu}; see \cite{Sarosi:2017ykf} for a nice review) has been the subject of intensive study in the past few years. 
For us, its main interesting feature is that it is an example of a UV completion of JT gravity, showing that JT gravity is a model of emergent gravity where the ideas mentioned in the introduction might naturally apply. 
In this section (that closely follows a section of \cite{Harlow:2018tqv}), I'll briefly review how this works. 

The SYK model is the QM of $N$ Majorana fermions with the Hamiltonian
\be \label{hsyk}
H_{int} = \frac 1{4!} \sum_{abcd} J_{abcd}\chi^a\chi^b\chi^c\chi^d\,,
\ee
where $J_{abcd}$ is drawn at random from a Gaussian ensemble with mean 0 and variance $J/N^{3/2}$. 
To do holography on a bulk interval, we'll need two copies of it, so the dynamical variables are 
$\chi^{a}_{i}$, $i \in L, R$. We'll also take the disorder to be the same on both sides.

To make it manifest that the model classicalizes at large $N$, it's convenient to exactly rewrite the partition function averaged over $J$,
\be
\langle Z\rangle_J = \int \mathcal{D}J\mathcal{D}\chi \exp\left[\sum_{i \in L,R} L_{i} - \frac{N^3}{12 J^2} (J_{abcd})^2 \right]\,,
\ee
in terms of ``master fields" $G$ and $\Sigma$:
\begin{eqnarray}\label{zaj}
\langle Z\rangle_J &=& \int \mathcal{D}G\mathcal{D}\Sigma e^{-iS(G,\Sigma)}\,, \\
\label{sykeff}
S(G,\Sigma) &=& -\frac {iN}2 \log \det (\delta_{ij}\partial_t - i\Sigma_{ij}) + \frac N2 \int dt\int dt' (\Sigma_{ij}G_{ij} + \frac i 4 J^2G^4)\,.
\end{eqnarray}
The variation of \eqref{sykeff} gives the eom's 
\be\label{sykeoms}
\Sigma_{ij} = i J^2 G^3_{ij}\,, \qquad \partial_t' - i\Sigma = G^{-1}\,,
\ee
that indeed are the Schwinger-Dyson equations at leading order in $N$. 

In the limit 
\be\label{irlimit}
N \gg \beta J \gg 1\,,
\ee
we can drop the time derivative in the second line of \eqref{sykeoms}. The remaining eom's then acquire a large reparametrization symmetry. This symmetry is spontaneously broken by solutions to the eom's so we can write down an effective action for the Goldstone modes,
\be\label{seff}
S_{eff} = \frac N J S_{Sch}\,,
\ee
where $S_{Sch}$ is short-hand for two copies of Schwarzian QM (whose action I'll omit here, see the references) with a diagonal $SL(2,R)$ gauged.
On the other hand, one finds the same action \eqref{seff} 
by integrating out the bulk in the JT theory \eqref{jtaction}, \cite{Maldacena:2016upp, Jensen:2016pah}. 

To summarize, JT gravity emerges from the SYK model in the low temperature limit  \eqref{irlimit}, and thus fits into the paradigm where we might expect a version of \eqref{lgtee} to apply. Having said this, I won't refer to the SYK model in the rest of this paper, instead working entirely in the IR.

\subsection{Area operator}

We find the analog of the area term \eqref{bhe} in JT gravity by the usual Euclidean argument. Namely, we evaluate the Euclidean path integral on the disk with circumference $\beta$ on the classical saddle, then plug into the thermodynamic formula, 
\be
Z[\beta] \approx e^{-S_{cl}}, \qquad S = (1 - \beta\partial_{\beta}) \log Z\,.
\ee 

Doing so, we find that 
\be\label{aop}
S_{BH} =  \frac{\phi_h}{4G_N} = \frac{8\pi^2\alpha\phi_b}{\beta}\,.
\ee
The first equality looks more like an ``area operator" (see also \cite{Engelsoy:2016xyb}). In the second, we use that $\phi_h = \phi_b r_s$ for $r_s$ the Schwarzschild radius of the wormhole solution, and $r_s = 2\pi/\beta$ on smooth classical solutions.

\subsection{Bulk formulation as a gauge theory}

The action \eqref{jtaction} can be rewritten as the action of a topological BF gauge theory with gauge group $SL(2,R)$ \cite{Isler:1989hq}, \cite{Chamseddine:1989yz}, 
\be\label{bfaction}
S_{BF} = \alpha \left(\int_M \Tr(BF)  + \frac 12 \int_{\partial M} \Tr (BA_0) \right)\,,
\ee
where $B^a$ is an adjoint scalar, $F$ is the two-form field strength of the gauge field $A$, and the trace is over the adjoint index. 
To see this, we form $A^a, B^a$ out of first-order gravity variables. This is the same idea as (indeed, a dimensional reduction of: see e.g. \cite{Mertens:2017mtv}, \cite{Mertens:2018fds}) the formulation of 3d pure gravity in $AdS_3$ as a $SL(2,R) \times SL(2,R)$ Chern-Simon theory \cite{Achucarro:1987vz}, \cite{Witten:1988hc}.
It is a change of variable in the bulk.

To make the boundary term vanish, we impose
\be\label{bcs}
\left. B^a\right|_{\partial M} = \left. 4\alpha \phi_b \sqrt{g_{00}} A^a_0\right|_{\partial M}
\ee
along the boundary. 
Since the bulk BF Hamiltonian equals the Gauss operator, all of the nontrivial dynamics come from the boundary part of the Hamiltonian, which depends on the constant of proportionality in \eqref{bcs}. Here I chose it so that the boundary Hamiltonian matches the one following from \eqref{jtaction}, \eqref{jtbc}. Since the coefficients are important in this paper I will do some checks below.

The bulk BF action is quite similar to the action of 2d Yang-Mills theory. We can use this to make a first pass at identifying its Hilbert space. Writing the bulk action in a $1+1d$ decomposition, 
\be
S_{BF} = \alpha \int dt\,\, dx (B^aD_0 A^a_1 - B^a D_1 A^a_0)\,,
\ee
we see that it has one pair of canonically conjugate variables $\{A^a_1, B^a\}$, subject to the Gauss law $D_x B^a(x) = 0$ imposed by $A^a_0$.
The Gauss law gives the same differential constraint on wavefunctions 
\be\label{e22}
\left(\partial_1 \frac{\delta}{\delta A_1^a(x)} + f^a_{bc} A^b_1(x) \frac{\delta}{\delta A^c_1(x)}\right)\Psi = 0
\ee
as in 2d Yang-Mills theory \cite{Cordes:1994fc}. Hence, if not for extra constraints coming from the boundary conditions, we could simply quote from \cite{Cordes:1994fc} that the Hilbert space of BF theory on ${\bf S}^1$ (for any gauge group) is the space of class functions of $G$, 
and the Hilbert space of the unconstrained BF theory on an interval is $L^2(G)$. These spaces are spanned by the characters and the matrix elements, respectively, of the reps of $G$ that appear in the Plancherel theorem. For $SL(2,R)$, these are the discrete and principal continuous reps.

However, the $AdS_2$ boundary conditions play a crucial role, truncating the Hilbert space relative to  $L^2(G)$. (See \cite{Blommaert:2018oro} for a very nice recent discussion, as well as all of the $SL(2,R)$ conventions that I am using in this note). \footnote{Other references that I found helpful while learning about $SL(2,R)$ are \cite{Kitaev:2017hnr}, \cite{Knapp}.}  
This is the 2d analog of the fact that pure gravity with $AdS_3$ boundary conditions is described not by an $SL(2,R)$ WZW edge theory but by a constrained theory due to the asymptotic behavior of the connection, that truncates the edge theory to Liouville (\cite{Coussaert:1995zp}; see e.g. \cite{Carlip:2005zn} for a review).

By dimensionally reducing the constrained 3d theory, one can show that the constrained BF description of JT gravity has the following features \cite{Blommaert:2018oro}:

\begin{itemize}\itemsep-0pt
\item[(*)] Its Hilbert space is spanned by a subset of matrix elements of the principal continuous reps only, namely matrix elements $|k\rangle = |k, \sqrt{\mu}, \sqrt{\mu}\rangle$ where the two indices denote eigenvalues of $J^-$ and $J^+$ (as opposed to the conventional situation where both are eigenvalues of $J^0$). 
\item[(*)] Such matrix elements are in the so-called mixed parabolic basis of $SL(2,R)$. The form of the matrix elements as functions over the $SL(2,R)$ group manifold is known explicitly. Interestingly, the Plancherel measure for these matrix elements is non-standard:
\be\label{psinh}
\dim k \rightarrow k \sinh 2\pi k\,.
\ee 
\end{itemize}

With this information, one can compute the disk partition function of the constrained BF theory using standard methods \cite{Blommaert:2018oro}. The partition function on a disk with circumference $\beta$ is 
\be\label{bfdisk}
Z_{{\bf D}^2}(\beta) = \int d\mu(k) e^{-\frac{\beta k^2}{4\alpha\phi_b}}
\ee
where $d\mu(k) =  k \sinh 2\pi k\, dk$, \eqref{psinh}. 
I will ignore the shift by $1/4$ in the Casimir here and below. 
\eqref{bfdisk} matches the (one-loop exact) partition function of the Schwarzian theory \cite{Stanford:2017thb} (see in particular their eq. (2.39).)  

I'll now construct the Hartle-Hawking state at inverse temperature $\beta$ in the BF theory.
The Hartle-Hawking state was recently constructed in gravity variables \cite{Harlow:2018tqv}. It is easier to construct in the gauge theory description.
Here the state at inverse temperature $\beta$ is set up by a path integral on the disk with length $\beta/2$, and with an open cut along a segment of the disk s.t. when we insert a state $|k\rangle$ in the cut, we get $\langle\Psi^{HH}_\beta|k\rangle$. 

Inserting a complete set of states in the Euclidean propagator and using that they are eigenfunctions of the Hamiltonian, 
\be
Z_{\bf D^2}(g,h, \beta/2) = \langle g|e^{- \beta H(k)/2}|h\rangle = \int dk \langle g|k\rangle \langle k|h\rangle e^{-\frac{\beta k^2}{8\alpha\phi_b}}\,.
\ee

Plugging in the analog of the identity element for $g$ in the mixed parabolic basis, again see \cite{Blommaert:2018oro},
\be\label{222}
Z_{\bf D^2}(g, \bone, \beta/2)= \int dk \sqrt{k \sinh 2\pi k} \, \langle g|k\rangle e^{-\frac{\beta k^2}{8\alpha\phi_b}}
\ee
where I've again dropped a $k$-independent normalization. 
Stripping off the $|k\rangle$ in \eqref{222}, the unnormalized Hartle-Hawking state is
\be\label{psihh}
|\Psi^{HH}_\beta\rangle = \sqrt{k\sinh 2\pi k}\,\, e^{-\frac{\beta k^2}{8\alpha\phi_b}}|k\rangle\,.
\ee
This can be compared to eq. (3.24) in \cite{Harlow:2018tqv}.
%

\section{Main calculation} \label{s3}

I'll now compute the entanglement entropy across the midpoint of the interval in the state $\Psi_\beta^{HH}(k)$ by using the Euclidean replica trick
\be\label{replica}
S = \left. -\frac{\partial}{\partial n}\frac{Z_n}{(Z_1)^n}\right|_{n=1}
\ee
for $Z_n$ the $n$-sheeted replica manifold, and deliberately writing everything in terms of the normalized probability distribution $p_{k,\beta}$, \eqref{hhprob}. This way of organizing the replica trick calculation was done in compact 2d Yang-Mills theory  in \cite{Donnelly:2014gva}. 

There is an important subtlety here. Whenever we do the replica trick, we should really excise a codimension-1 tubular region around the conical singularity and put boundary conditions there  (e.g. \cite{Ohmori:2014eia}). This problem is especially acute in gauge theories where the Hilbert space doesn't factorize and different prescriptions affect both the universal and non-universal data (see \cite{Lin:2018bud} and refs therein). Here I will assume that our boundary condition is such that the excised region can be shrunk to a point and removed without changing the naive topology of the replicated manifold. One justification for this is that in Chern-Simons theory with a compact gauge group, the shrinkable boundary condition is known to correctly reproduce the topological EE \cite{Dong:2008ft}, \cite{jafferis}. Another is that the BF theory on smooth manifolds is equivalent to the Schwarzian theory on the boundary, so if we assume that we can do the replica trick on smooth manifolds, we will just be rewriting the Schwarzian entropy in a suggestive manner.

With this assumption, the $n$-replicated manifold still has the topology of a disk. The only difference from \eqref{bfdisk} is that it has a $n$-times-longer circumference,
\be
Z_n = \int d\mu(k) e^{-\frac{n \beta k^2}{4\alpha\phi_b}}\,.
\ee
We rewrite it as
\be
Z_n = \int d\mu(k)^{1-n}(Z_1\, p_{k,\beta})^n
\ee
where
\be\label{hhprob}
p_{k, \beta} = Z_1^{-1}|\langle \Psi^{HH}_\beta |k\rangle|^2 = Z_1^{-1}\,\,k \sinh 2\pi k \,e^{-\frac{\beta k^2}{4\alpha\phi_b}}
\ee
is the normalized probability distribution of the Hartle-Hawking state.
Plugging into \eqref{replica}, we immediately find  
\be\label{result}
S_{EE}(\Psi^{HH}_\beta) = \int_k p_k[-\log p_k + \log k \sinh 2\pi k]\,.
\ee

I now show that the second term agrees with the JT analog of the Bekenstein-Hawking entropy, \eqref{aop}, in the classical (large $\alpha$) limit. The Hartle-Hawking wavefunction is peaked at 
\be\label{kmax}
k = \frac{4\pi\alpha\phi_b}{\beta} 
\ee
at leading order in $\alpha$. Evaluating the $\log \dim R$-type edge term at the peak of the wavefunction, 
\be\label{matchee}
S_{EE}(\Psi^{HH}_\beta) \left. = \log \sinh 2\pi k\right|_{k = 4\pi \alpha \phi_b/\beta } = \frac{8\pi^2\alpha \phi_b}{\beta} = \frac{\langle\phi_h\rangle}{4G_N}\,,
\ee
in agreement with \eqref{aop}.
The first term gives a correction but it is subleading at large $k$.
This demonstrates a version of \eqref{conjecture} in JT gravity. 

If we view ``$\sinh 2\pi k"$ as an element of the modular S-matrix of Liouville CFT, \eqref{matchee} is essentially the same formula as the (dimensional reduction of the) main result of \cite{Mcgough:2013gka}. \eqref{result}, \eqref{matchee} gives an explanation and the form of the subleading corrections.

On a technical level, \eqref{matchee} had to work (other than the corrections necessarily being subleading) since as mentioned above, the constrained BF theory on a smooth manifold is equivalent to the Schwarzian theory on the boundary, and the Schwarzian theory is known to reproduce the correct entropy \cite{Maldacena:2016upp}.
The main benefit that we get from rewriting the Schwarzian thermal entropy in the form \eqref{result} is that we can now use our knowledge of entanglement entropy in compact gauge theories to interpret what we are counting. 

\section{Interpretation}\label{s4}


In the literature on entanglement entropy in compact gauge theories (see appendix \ref{a1} for a brief review), a $\log \dim R$-type entanglement edge term has appeared in two (related) situations.
In an emergent gauge theory, it arises as a boundary entropy counting the correlations between UV degrees of freedom at the entangling surface which are the microscopic constituents of a Wilson loop in rep $R$ (e.g. \cite{Harlow:2015lma}). Alternatively, when one inserts a background Wilson line in rep $R$ ending on static charges in a Chern-Simons theory, the $\log \dim R$ term is more naturally viewed as counting the entanglement between the charges, which may be located far from the entangling surface.
\footnote{In the literature, the extra entanglement from inserting a background Wilson line in Chern-Simons theory was also computed by adding UV degrees of freedom at the entangling surface  \cite{Kitaev:2005dm, Fliss:2017wop, Wong:2017pdm}, but this will not always work when the gauge group is noncompact.}

In both cases, the $\log \dim R$ edge term has the following properties:
\begin{itemize}\itemsep-0pt
\item[(*)] it quantifies entanglement that cannot be assigned to a gauge-invariant operator algebra represented by the IR Hilbert space. From the IR point of view, it is more like a label on the states. 
(The Shannon-like term, which is the subleading correction in \eqref{result}, captures the EE of gauge-invariant operators in the subregion).
\footnote{More precisely, in a compact $1+1d$ gauge theory, the Shannon term $\sum - p_k \log p_k$ captures the algebraic EE of the gauge-invariant operators contained in a region which are the Casimirs, see appendix \ref{a1}. In this example, since the Hilbert space has a continuous spectrum, the Shannon term became a differential entropy in \eqref{result} while the algebraic EE is not well-defined. I think that this term must be the algebraic EE with any reasonable regulator, though this should perhaps be checked explicitly. The point is that in a $1+1d$ gauge theory, the separation of terms appearing in the replica trick into Shannon-like and $``\log \dim R"$ terms is a UV/IR separation at the scale of the emergence of the gauge theory, \eqref{irlimit}.}
\item[(*)] That label refers to the representation of a non-dynamical Wilson line, and the $\log \dim R$ term can be interpreted as the EE of a pair of maximally entangled static charges in that rep.
\end{itemize}


Generalizing to a noncompact gauge group, it seems natural to view the state $|k\rangle$ of the constrained BF theory as implicitly containing a non-dynamical Wilson line in the principal continuous rep $k$ of $SL(2,R)$, and to assign the  $\log k \sinh 2 \pi k $ edge term in the EE across the interval to this Wilson line. Let me remain agnostic about where the microscopic dof's are located for now, I will come back to this shortly. From \eqref{matchee}, the proposal is then that the (two-sided) black hole entropy is counting states in the continuous reps of $SL(2,R)$ (although I have not yet explained which states these are supposed to be).

Note that Wilson lines in the continuous reps of $SL(2,R)$ are not usually thought of as being contained in the algebra of the BF theory (e.g. \cite{Blommaert:2018oro}, \cite{Seiberg:1990eb}), so the black hole entropy is not an entanglement entropy from a totally IR point of view, consistent with the non-factorizability of the JT Hilbert space. In other words, the JT theory has no black hole microstates \cite{Harlow:2018tqv} so we have to ``extend the Hilbert space" if we want to count the black hole entropy. In this picture, the Hilbert space of JT gravity is really a union of single-state superselection sectors, each being the analog of a Chern-Simons Hilbert space in the presence of a different set of sources, and the black hole entropy is topological EE \cite{Mcgough:2013gka}. 


The assignation of the entropy to states in the continuous reps of $SL(2,R)$ is consistent with some work from previous decades on the entropy of the BTZ black hole (see \cite{Carlip:2005zn} for a review of this subject). Ref. \cite{Harlow:2018tqv} explained that 3d pure gravity should probably be thought of as being in the same universality class as JT gravity, i.e. as a theory of emergent gravity that contains wormholes but not black hole microstates. 
In the past, people attempted to find the microstates of the BTZ black hole in the Liouville edge CFT of 3d pure gravity, but the normalizable sector of Liouville CFT doesn't have nearly enough states to account for the black hole entropy \cite{Kutasov:1990sv}, \cite{Martinec:1998wm}. However, Chen was able to reproduce the entropy by counting some states in the non-normalizable sector \cite{Chen:2004rh} (although with several technical puzzles left open). This work seems related to the picture here, though the details are quite different.

\subsection{Where are the microstates located?}\label{s41}

In a gauge theory with a compact gauge group, the $\log \dim R$ edge term can always be thought of as a boundary entropy. Essentially this is because we can always cut a Wilson line in a compact gauge theory by adding charges at the entangling surface.
The most naive extension of this interpretation to gravity is to speculate that black hole entropy can be understood from the number of ways to glue two patches of space together at a geometric interface. This was the interpretation that I suggested earlier \cite{Lin:2017uzr} by the analogy to the compact gauge theory; previously it had been suggested and studied at the classical level in \cite{Donnelly:2016auv} (see also \cite{Speranza:2017gxd}), and it also seems similar in spirit to a much earlier calculation by Carlip \cite{Carlip:1994gy} (see also \cite{Maldacena:1998ih}).

We can study this idea here by comparing the above replica trick calculation for the $SL(2,R)$ BF theory on the interval  with the same calculation for the $SL(2,R)$ BF theory on spatial ${\bf S}^1$. \footnote{ See \cite{Freidel:2002xb}, \cite{Constantinidis:2008ty} for more about JT gravity on ${\bf S}^1$. } The operation of forcing a gauge theory to factorize at a geometric boundary by adding a minimal number of pure gauge degrees of freedom at the boundary, which I will call the ``extended Hilbert space construction" (see appendix \ref{a1}), is local and shouldn't care about the asymptotic boundary conditions. However, in the replica trick calculation  on ${\bf S}^1$, we find the replacement rule 
\be
\dim k \rightarrow k \tanh \pi k
\ee
(i.e. the standard Plancherel measure for $SL(2,R)$), instead of \eqref{psinh}.
This does not grow fast enough at large $k$ to account for black hole entropy.

Hence, the extended Hilbert space construction does not seem relevant for explaining the area term. (See \cite{Donnelly:2018nbv} for recent criticism). Perhaps there is still some way to interpret the wormhole entropy in a theory of emergent gravity by extending the theory with charges that let us cut the gravitational Wilson lines, 
but it should (maybe unsurprisingly, in retrospect) be less local. 

Perhaps a better cartoon of the bulk is to imagine that the JT theory comes with asymptotic dof's in the continuous reps of $SL(2)$ that are infinitely massive in the limit $G\hbar \rightarrow 0$, whose entanglement gives rise to the bulk.
Of course, more work is needed to make this precise.

\section{Discussion}\label{s5}

To summarize, the black hole entropy associated to an $AdS_2$ wormhole in $1+1d$ JT gravity can be matched to a $\log \dim R$ type entanglement edge term in the BF formulation with exact agreement of the coefficient $1/4G_N$, \eqref{result2}. This has some interesting implications. One is that at least in this example, the identification of the RT formula \eqref{rt} with the formula for EE in an emergent gauge theory \eqref{lgtee} seems to be true, the meaning of the RT formula is ``boundary EE = bulk EE" when one includes the edge modes, and the
explanation behind the universality of the area term is that it is related to a natural measure on the symmetry group. These things should be understood more generally. 

Using our knowledge of EE in compact gauge theories, I then commented on what we might be counting. 
Somewhat fancifully, our picture is that ``the bulk emerges from the entanglement of effective atoms of space" which here are SL(2)-charged heavy sources in the continuous reps. 
However, this part is speculative in the absence of an explicit construction of the would-be microstates. 
More precisely, in a compact gauge theory it is very clear what the $\log \dim R$ entanglement edge term is counting: there a pair of background charges in rep $R$ are maximally entangled, and the EE just counts the size of the matrix. But if we want to view the noncompact edge term similarly, we have to explain why the Plancherel measure of $PSL(2,R)$ provides a physical regulator on the trace of an infinite-dimensional rep. 



There are many possibilities for future work. Here are a few of them:

It might be illuminating to translate everything back to the gravity variables.

It would be interesting to generalize this argument to pure 3d quantum gravity (where there are many related results, especially \cite{Mcgough:2013gka}, also \cite{Ammon:2013hba}, \cite{deBoer:2013vca} and the follow-ups). Then we can see what happens when the horizon has a finite area.
Generalizing to $ d \geq 4$ seems harder since there is no decoupled pure gravity sector. 

If these ideas are correct, it would be interesting to understand the implications for the outstanding problems in black hole physics. 

In the interpretation, the static charges would become dynamical in a UV completion of gravity. It would be interesting to understand how they arise in the SYK model (where the heavy modes indeed appear to organize into the above-mentioned reps of $SL(2)$, e.g. \cite{Kitaev:2017awl}) and also in string theory. Relatedly, maybe we can understand the factor of $N$ in the entanglement entropy across a boundary-anchored string \cite{Susskind:1994sm}, \cite{Lewkowycz:2013laa}.  

\subsection*{Acknowledgments}
I'd like to thank Daniel Harlow, Matt Headrick, Juan Maldacena, Eric Mintun, Djordje Radicevic, Mukund Rangamani, David Simmons-Duffin and Bogdan Stoica for helpful comments and discussions. I'd also like to thank the Perimeter Institute and Galileo Galilei Institute for hospitality as this work was being completed. My work is supported by the William D. Loughlin Membership at the IAS and by the U.S. Department of Energy.

\begin{appendix}
\section{A brief introduction to EE in gauge theories}\label{a1}
This section contains a very short introduction to entanglement entropy in gauge theories with a compact gauge group, and especially the $\log \dim R$ term in the extended Hilbert space construction, that I'll define below. It's meant for readers who haven't previously looked into EE in gauge theories. See section 8 of \cite{Lin:2018bud} for a more thorough review.
As mentioned in the main text, I don't think that the details of the extended Hilbert space construction are relevant for gravity. (More precisely, although the $\log \dim R$ term in the main text can also presumably be assigned to {\it an} extension of the Hilbert space, it doesn't come from locally minimally extending the Hilbert space at the entangling cut, as explained in section \ref{s41}). However, the following example is still the fastest way to get acquainted with {\it an} example of a $\log \dim R$ term, and to get some intuition for what it counts. 

Let's start by reviewing the context for the problem. To assign an entanglement entropy to a region $A$ in a state $|\Psi\rangle$ of a quantum system, we usually assume that the Hilbert space factorizes, $\HH = \HH_A \otimes \HH_{\bar A}\,.$ We then assign a density matrix $$\rho_A = \Tr_{\bar A}|\Psi\rangle\langle\Psi|$$ to region $A$ by taking a partial trace, and take its von Neumann entropy, 
\be\label{defee}
S_{EE}(\rho_A) = -\Tr_A \rho_A \log \rho_A\,,
\ee
to be the entanglement entropy.

Continuum QFT's never have a factorizable Hilbert space, e.g. because the short-distance structure is universal in all the states of the QFT. What people often do is to implicitly regulate the QFT with a lattice, yielding a clear-cut factorization for QFT's whose fundamental dof's are local (as e.g. in the earliest calculations \cite{Srednicki:1993im}). But this is not enough when the QFT is a gauge theory because of Gauss's law.

To assign a density matrix $\rho_A$ nonetheless to a subset of links in a lattice gauge theory, two inequivalent definitions were proposed in the past few years:

\begin{enumerate}
\item In the algebraic approach (e.g. \cite{Casini:2013rba}), we assign a density operator and entanglement entropy to a(ny) gauge-invariant subalgebra as follows. Given a subalgebra $\AAA_0$, (for finite-dimensional systems) there is in general a unique element $\rho_{\AAA_0} \in \AAA_0$ that correctly reproduces the vev's of all the other operators in $\AAA_0$ when used as a density operator. We take its von Neumann entropy to be the EE. \footnote{To complete the definition, we also have to pick a Hilbert space to take the trace over, but different choices will just differ by a constant related to the relative sizes of the Hilbert spaces. There's an algorithm that one can follow to get an answer that agrees with \eqref{defee} when the Hilbert space factorizes \cite{Casini:2013rba}.}

An especially natural choice, which is what I call $S_{alg}(A)$ in the main text (and which is called the electric center choice elsewhere), is to take the EE of a region $A$ to be the algebraic EE of the maximal gauge-invariant operator algebra fully supported on the region.
\item In the extended Hilbert space construction (e.g. \cite{Buividovich:2008gq}, \cite{Donnelly:2011hn}, \cite{Ghosh:2015iwa}), we embed the Hilbert space of the lattice gauge theory in the minimal larger one that factorizes across $\partial A$,
\be\label{hext}
\HH_{phys} \subset \HH_{ext.} = \HH_A \otimes \HH_{\bar{A}}\,, 
\ee
by lifting the Gauss constraint at the boundary sites. We can then define 
\be
\rho_A = \Tr_{\bar A, \HH_{ext.}}\rho\,,
\ee
and take its von Neumann entropy to define the EE.
\end{enumerate}

The extended Hilbert space construction is the fastest way to get a first look at a $``\log \dim R$"-type edge term. Let's take it for a test run in the simplest lattice gauge theory, which is a lattice with two links and two nodes. 
I.e. we compute the EE across an interval in a 2d Yang-Mills theory (with a compact gauge group) on ${\bf S}^1$. 
A nice reference for this example is \cite{Donnelly:2014gva}.
Here both of the nodes are boundary nodes.
We are instructed to construct $\HH_{ext.}$ by lifting the Gauss law at both of them, so $\HH_{ext.}$ is the tensor product of two copies of the Hilbert space of the gauge theory on an interval. 

The Hilbert space of 2d Yang-Mills on ${\bf S}^1$ is the space of class functions, and the Hilbert space of 2d Yang-Mills on an interval is $L^2(G)$. It is well-known that bases for these spaces are furnished by the group characters and the matrix elements, respectively.
 The embedding $\HH_{phys} \subset \HH_{ext.}$ follows from doubling the trace definition of the group character. With the correctly normalized basis elements
$
|R\rangle = \chi_R(g)\,,$ on ${\bf S}^1$ and 
$|R,i,j\rangle = \sqrt{\dim R}\,\,U_{ij}^R(g)$ on the interval, 
\begin{eqnarray}
|R\rangle &=& (\dim R)^{-1/2}\sum_{i \in 1, \dots, R} |R,i,i\rangle \\
&=& (\dim R)^{-1}\sum_{i,j\in 1, \dots, R}|R,i,j\rangle \otimes |R,j,i\rangle\,,\label{defchar}
\end{eqnarray}
where the first line is the definition of the character and the second repeats it.

From \eqref{defchar}, we can compute the EE in the most general state $\Psi(R)|R\rangle \in \HH_{phys}$.
With
\be
\Tr_{\bar A}|R\rangle \langle R| = (\dim R)^{-2} \sum_{i,j} |R,i,j\rangle\langle R,i,j|_A\,,
\ee
we find that 
\be\label{2dee}
S_{EE} = -\Tr\rho_A \log \rho_A = \sum_R p_R \left[- \log p_R + 2 \log \dim R\right]
\ee
for $p_R = |\Psi(R)|^2$\,.

It was pointed out in \cite{Soni:2015yga} that the extended Hilbert space result for the EE, \eqref{2dee}, differs from the algebraic EE of maximal gauge-invariant operators in region $A$ by the $\log \dim R$ term. \footnote{I.e. algebraic EE of gauge-invariant operators is captured by the Shannon entropy only in a $1+1d$ pure gauge theory, that measures the correlations between the Casimirs mandated by Gauss's law, as mentioned above. In theories with more degrees of freedom, there would be other terms. See \eqref{lgtee}, \eqref{lgtee2}.} This is not surprising for the following reason. When we extended the Hilbert space in the step \eqref{hext}, we implicitly extended our operator algebra by adding to it Wilson lines in all representations ending on infinitely massive surface charges at the entangling surface (i.e., $\HH_{ext.}$ faithfully represents this larger algebra). From this perspective, {\it (i)} the $\log \dim R$ term is counting the correlations of the static charges, and {\it (ii)} we would not expect to see it in a definition of entanglement that only refers to gauge-invariant operators in the IR. In this setting, this explains the two general features mentioned on page 8.

The extended Hilbert space construction seemed rather arbitrary from the point of view of the gauge theory. In \cite{Lin:2017uzr}, I pointed out that it gives the UV-exact EE (up to a state-independent constant) when a (compact) gauge theory is emergent. This motivated \eqref{conjecture} when viewing gravity in the bulk as an emergent gauge theory. In making this conjecture I applied \eqref{lgtee} outside its regime of validity, and as explained above, the details of the extended Hilbert space construction are probably too local for gravity. On the other hand, something like \eqref{conjecture} appears to actually be true.
\end{appendix}

\newpage
\bibliographystyle{ssg}
\bibliography{2djt}

\begingroup\raggedright\begin{thebibliography}{10}

\bibitem{Ryu:2006bv}
S.~Ryu and T.~Takayanagi, ``{Holographic derivation of entanglement entropy
  from AdS/CFT},'' {\em Phys. Rev. Lett.} {\bf 96} (2006) 181602,
  \href{http://xxx.lanl.gov/abs/hep-th/0603001}{{\tt hep-th/0603001}}.

\bibitem{Ryu:2006ef}
S.~Ryu and T.~Takayanagi, ``{Aspects of Holographic Entanglement Entropy},''
  {\em JHEP} {\bf 08} (2006) 045,
  \href{http://xxx.lanl.gov/abs/hep-th/0605073}{{\tt hep-th/0605073}}.

\bibitem{Faulkner:2013ana}
T.~Faulkner, A.~Lewkowycz, and J.~Maldacena, ``{Quantum corrections to
  holographic entanglement entropy},'' {\em JHEP} {\bf 11} (2013) 074,
  \href{http://xxx.lanl.gov/abs/1307.2892}{{\tt 1307.2892}}.

\bibitem{Dong:2016eik}
X.~Dong, D.~Harlow, and A.~C. Wall, ``{Reconstruction of Bulk Operators within
  the Entanglement Wedge in Gauge-Gravity Duality},'' {\em Phys. Rev. Lett.}
  {\bf 117} (2016), no.~2 021601,
  \href{http://xxx.lanl.gov/abs/1601.05416}{{\tt 1601.05416}}.

\bibitem{Harlow:2016vwg}
D.~Harlow, ``{The Ryu{\^a}€``Takayanagi Formula from Quantum Error
  Correction},'' {\em Commun. Math. Phys.} {\bf 354} (2017), no.~3 865--912,
  \href{http://xxx.lanl.gov/abs/1607.03901}{{\tt 1607.03901}}.

\bibitem{Lin:2017uzr}
J.~Lin, ``{Ryu-Takayanagi Area as an Entanglement Edge Term},''
  \href{http://xxx.lanl.gov/abs/1704.07763}{{\tt 1704.07763}}.

\bibitem{Donnelly:2016auv}
W.~Donnelly and L.~Freidel, ``{Local subsystems in gauge theory and gravity},''
  {\em JHEP} {\bf 09} (2016) 102,
  \href{http://xxx.lanl.gov/abs/1601.04744}{{\tt 1601.04744}}.

\bibitem{Buividovich:2008gq}
P.~V. Buividovich and M.~I. Polikarpov, ``{Entanglement entropy in gauge
  theories and the holographic principle for electric strings},'' {\em Phys.
  Lett.} {\bf B670} (2008) 141--145,
  \href{http://xxx.lanl.gov/abs/0806.3376}{{\tt 0806.3376}}.

\bibitem{Donnelly:2011hn}
W.~Donnelly, ``{Decomposition of entanglement entropy in lattice gauge
  theory},'' {\em Phys. Rev.} {\bf D85} (2012) 085004,
  \href{http://xxx.lanl.gov/abs/1109.0036}{{\tt 1109.0036}}.

\bibitem{Ghosh:2015iwa}
S.~Ghosh, R.~M. Soni, and S.~P. Trivedi, ``{On The Entanglement Entropy For
  Gauge Theories},'' {\em JHEP} {\bf 09} (2015) 069,
  \href{http://xxx.lanl.gov/abs/1501.02593}{{\tt 1501.02593}}.

\bibitem{Radicevic:2015sza}
o.~Radievi, ``{Entanglement in Weakly Coupled Lattice Gauge Theories},'' {\em
  JHEP} {\bf 04} (2016) 163, \href{http://xxx.lanl.gov/abs/1509.08478}{{\tt
  1509.08478}}.

\bibitem{Harlow:2018tqv}
D.~Harlow and D.~Jafferis, ``{The Factorization Problem in Jackiw-Teitelboim
  Gravity},'' \href{http://xxx.lanl.gov/abs/1804.01081}{{\tt 1804.01081}}.

\bibitem{Jackiw:1984je}
R.~Jackiw, ``{Lower Dimensional Gravity},'' {\em Nucl. Phys.} {\bf B252} (1985)
  343--356.

\bibitem{Teitelboim:1983ux}
C.~Teitelboim, ``{Gravitation and Hamiltonian Structure in Two Space-Time
  Dimensions},'' {\em Phys. Lett.} {\bf 126B} (1983) 41--45.

\bibitem{Sachdev:1992fk}
S.~Sachdev and J.~Ye, ``{Gapless spin fluid ground state in a random, quantum
  Heisenberg magnet},'' {\em Phys. Rev. Lett.} {\bf 70} (1993) 3339,
  \href{http://xxx.lanl.gov/abs/cond-mat/9212030}{{\tt cond-mat/9212030}}.

\bibitem{Kitaev}
A.~Kitaev, ``A simple model of quantum holography.'' Talks at KITP 2015.

\bibitem{Maldacena:2016hyu}
J.~Maldacena and D.~Stanford, ``{Remarks on the Sachdev-Ye-Kitaev model},''
  {\em Phys. Rev.} {\bf D94} (2016), no.~10 106002,
  \href{http://xxx.lanl.gov/abs/1604.07818}{{\tt 1604.07818}}.

\bibitem{Sarosi:2017ykf}
G.~Srosi, ``{AdS$_{2}$ holography and the SYK model},'' {\em PoS} {\bf
  Modave2017} (2018) 001, \href{http://xxx.lanl.gov/abs/1711.08482}{{\tt
  1711.08482}}.

\bibitem{Maldacena:2016upp}
J.~Maldacena, D.~Stanford, and Z.~Yang, ``{Conformal symmetry and its breaking
  in two dimensional Nearly Anti-de-Sitter space},'' {\em PTEP} {\bf 2016}
  (2016), no.~12 12C104, \href{http://xxx.lanl.gov/abs/1606.01857}{{\tt
  1606.01857}}.

\bibitem{Jensen:2016pah}
K.~Jensen, ``{Chaos in AdS$_2$ Holography},'' {\em Phys. Rev. Lett.} {\bf 117}
  (2016), no.~11 111601, \href{http://xxx.lanl.gov/abs/1605.06098}{{\tt
  1605.06098}}.

\bibitem{Engelsoy:2016xyb}
J.~Engelsy, T.~G. Mertens, and H.~Verlinde, ``{An investigation of AdS$_{2}$
  backreaction and holography},'' {\em JHEP} {\bf 07} (2016) 139,
  \href{http://xxx.lanl.gov/abs/1606.03438}{{\tt 1606.03438}}.

\bibitem{Isler:1989hq}
K.~Isler and C.~A. Trugenberger, ``{A Gauge Theory of Two-dimensional Quantum
  Gravity},'' {\em Phys. Rev. Lett.} {\bf 63} (1989) 834.

\bibitem{Chamseddine:1989yz}
A.~H. Chamseddine and D.~Wyler, ``{Gauge Theory of Topological Gravity in
  (1+1)-Dimensions},'' {\em Phys. Lett.} {\bf B228} (1989) 75--78.

\bibitem{Mertens:2017mtv}
T.~G. Mertens, G.~J. Turiaci, and H.~L. Verlinde, ``{Solving the Schwarzian via
  the Conformal Bootstrap},'' {\em JHEP} {\bf 08} (2017) 136,
  \href{http://xxx.lanl.gov/abs/1705.08408}{{\tt 1705.08408}}.

\bibitem{Mertens:2018fds}
T.~G. Mertens, ``{The Schwarzian Theory - Origins},'' {\em JHEP} {\bf 05}
  (2018) 036, \href{http://xxx.lanl.gov/abs/1801.09605}{{\tt 1801.09605}}.

\bibitem{Achucarro:1987vz}
A.~Achucarro and P.~K. Townsend, ``{A Chern-Simons Action for Three-Dimensional
  anti-De Sitter Supergravity Theories},'' {\em Phys. Lett.} {\bf B180} (1986)
  89. [,732(1987)].

\bibitem{Witten:1988hc}
E.~Witten, ``{(2+1)-Dimensional Gravity as an Exactly Soluble System},'' {\em
  Nucl. Phys.} {\bf B311} (1988) 46.

\bibitem{Cordes:1994fc}
S.~Cordes, G.~W. Moore, and S.~Ramgoolam, ``{Lectures on 2-d Yang-Mills theory,
  equivariant cohomology and topological field theories},'' {\em Nucl. Phys.
  Proc. Suppl.} {\bf 41} (1995) 184--244,
  \href{http://xxx.lanl.gov/abs/hep-th/9411210}{{\tt hep-th/9411210}}.

\bibitem{Blommaert:2018oro}
A.~Blommaert, T.~G. Mertens, and H.~Verschelde, ``{The Schwarzian Theory - A
  Wilson Line Perspective},'' \href{http://xxx.lanl.gov/abs/1806.07765}{{\tt
  1806.07765}}.

\bibitem{Kitaev:2017hnr}
A.~Kitaev, ``{Notes on $\widetilde{\mathrm{SL}}(2,\mathbb{R})$
  representations},'' \href{http://xxx.lanl.gov/abs/1711.08169}{{\tt
  1711.08169}}.

\bibitem{Knapp}
A.~Knapp, {\em Representation Theory of Semisimple Groups}.
\newblock Princeton University Press, 1986.

\bibitem{Coussaert:1995zp}
O.~Coussaert, M.~Henneaux, and P.~van Driel, ``{The Asymptotic dynamics of
  three-dimensional Einstein gravity with a negative cosmological constant},''
  {\em Class. Quant. Grav.} {\bf 12} (1995) 2961--2966,
  \href{http://xxx.lanl.gov/abs/gr-qc/9506019}{{\tt gr-qc/9506019}}.

\bibitem{Carlip:2005zn}
S.~Carlip, ``{Conformal field theory, (2+1)-dimensional gravity, and the BTZ
  black hole},'' {\em Class. Quant. Grav.} {\bf 22} (2005) R85--R124,
  \href{http://xxx.lanl.gov/abs/gr-qc/0503022}{{\tt gr-qc/0503022}}.

\bibitem{Stanford:2017thb}
D.~Stanford and E.~Witten, ``{Fermionic Localization of the Schwarzian
  Theory},'' {\em JHEP} {\bf 10} (2017) 008,
  \href{http://xxx.lanl.gov/abs/1703.04612}{{\tt 1703.04612}}.

\bibitem{Donnelly:2014gva}
W.~Donnelly, ``{Entanglement entropy and nonabelian gauge symmetry},'' {\em
  Class. Quant. Grav.} {\bf 31} (2014), no.~21 214003,
  \href{http://xxx.lanl.gov/abs/1406.7304}{{\tt 1406.7304}}.

\bibitem{Ohmori:2014eia}
K.~Ohmori and Y.~Tachikawa, ``{Physics at the entangling surface},'' {\em J.
  Stat. Mech.} {\bf 1504} (2015) P04010,
  \href{http://xxx.lanl.gov/abs/1406.4167}{{\tt 1406.4167}}.

\bibitem{Lin:2018bud}
J.~Lin and D.~Radicevic, ``{Comments on Defining Entanglement Entropy},''
  \href{http://xxx.lanl.gov/abs/1808.05939}{{\tt 1808.05939}}.

\bibitem{Dong:2008ft}
S.~Dong, E.~Fradkin, R.~G. Leigh, and S.~Nowling, ``{Topological Entanglement
  Entropy in Chern-Simons Theories and Quantum Hall Fluids},'' {\em JHEP} {\bf
  05} (2008) 016, \href{http://xxx.lanl.gov/abs/0802.3231}{{\tt 0802.3231}}.

\bibitem{jafferis}
D.~Jafferis, ``Factorization in gravity.'' Talk at Bariloche meeting.

\bibitem{Mcgough:2013gka}
L.~McGough and H.~Verlinde, ``{Bekenstein-Hawking Entropy as Topological
  Entanglement Entropy},'' {\em JHEP} {\bf 11} (2013) 208,
  \href{http://xxx.lanl.gov/abs/1308.2342}{{\tt 1308.2342}}.

\bibitem{Harlow:2015lma}
D.~Harlow, ``{Wormholes, Emergent Gauge Fields, and the Weak Gravity
  Conjecture},'' {\em JHEP} {\bf 01} (2016) 122,
  \href{http://xxx.lanl.gov/abs/1510.07911}{{\tt 1510.07911}}.

\bibitem{Kitaev:2005dm}
A.~Kitaev and J.~Preskill, ``{Topological entanglement entropy},'' {\em Phys.
  Rev. Lett.} {\bf 96} (2006) 110404,
  \href{http://xxx.lanl.gov/abs/hep-th/0510092}{{\tt hep-th/0510092}}.

\bibitem{Fliss:2017wop}
J.~R. Fliss, X.~Wen, O.~Parrikar, C.-T. Hsieh, B.~Han, T.~L. Hughes, and R.~G.
  Leigh, ``{Interface Contributions to Topological Entanglement in Abelian
  Chern-Simons Theory},'' {\em JHEP} {\bf 09} (2017) 056,
  \href{http://xxx.lanl.gov/abs/1705.09611}{{\tt 1705.09611}}.

\bibitem{Wong:2017pdm}
G.~Wong, ``{A note on entanglement edge modes in Chern Simons theory},''
  \href{http://xxx.lanl.gov/abs/1706.04666}{{\tt 1706.04666}}.

\bibitem{Seiberg:1990eb}
N.~Seiberg, ``{Notes on quantum Liouville theory and quantum gravity},'' {\em
  Prog. Theor. Phys. Suppl.} {\bf 102} (1990) 319--349.

\bibitem{Kutasov:1990sv}
D.~Kutasov and N.~Seiberg, ``{Number of degrees of freedom, density of states
  and tachyons in string theory and CFT},'' {\em Nucl. Phys.} {\bf B358} (1991)
  600--618.

\bibitem{Martinec:1998wm}
E.~J. Martinec, ``{Conformal field theory, geometry, and entropy},''
  \href{http://xxx.lanl.gov/abs/hep-th/9809021}{{\tt hep-th/9809021}}.

\bibitem{Chen:2004rh}
Y.~Chen, {\em {Quantum Liouville theory and BTZ black hole entropy}}.
\newblock PhD thesis, UC, Davis, 2004.

\bibitem{Speranza:2017gxd}
A.~J. Speranza, ``{Local phase space and edge modes for
  diffeomorphism-invariant theories},'' {\em JHEP} {\bf 02} (2018) 021,
  \href{http://xxx.lanl.gov/abs/1706.05061}{{\tt 1706.05061}}.

\bibitem{Carlip:1994gy}
S.~Carlip, ``{The Statistical mechanics of the (2+1)-dimensional black hole},''
  {\em Phys. Rev.} {\bf D51} (1995) 632--637,
  \href{http://xxx.lanl.gov/abs/gr-qc/9409052}{{\tt gr-qc/9409052}}.

\bibitem{Maldacena:1998ih}
J.~M. Maldacena and A.~Strominger, ``{Statistical entropy of de Sitter
  space},'' {\em JHEP} {\bf 02} (1998) 014,
  \href{http://xxx.lanl.gov/abs/gr-qc/9801096}{{\tt gr-qc/9801096}}.

\bibitem{Freidel:2002xb}
L.~Freidel and E.~R. Livine, ``{Spin networks for noncompact groups},'' {\em J.
  Math. Phys.} {\bf 44} (2003) 1322--1356,
  \href{http://xxx.lanl.gov/abs/hep-th/0205268}{{\tt hep-th/0205268}}.

\bibitem{Constantinidis:2008ty}
C.~P. Constantinidis, O.~Piguet, and A.~Perez, ``{Quantization of the
  Jackiw-Teitelboim model},'' {\em Phys. Rev.} {\bf D79} (2009) 084007,
  \href{http://xxx.lanl.gov/abs/0812.0577}{{\tt 0812.0577}}.

\bibitem{Donnelly:2018nbv}
W.~Donnelly and S.~B. Giddings, ``{Gravitational splitting at first-order:
  quantum information localization in gravity},''
  \href{http://xxx.lanl.gov/abs/1805.11095}{{\tt 1805.11095}}.

\bibitem{Ammon:2013hba}
M.~Ammon, A.~Castro, and N.~Iqbal, ``{Wilson Lines and Entanglement Entropy in
  Higher Spin Gravity},'' {\em JHEP} {\bf 10} (2013) 110,
  \href{http://xxx.lanl.gov/abs/1306.4338}{{\tt 1306.4338}}.

\bibitem{deBoer:2013vca}
J.~de~Boer and J.~I. Jottar, ``{Entanglement Entropy and Higher Spin Holography
  in AdS$_3$},'' {\em JHEP} {\bf 04} (2014) 089,
  \href{http://xxx.lanl.gov/abs/1306.4347}{{\tt 1306.4347}}.

\bibitem{Kitaev:2017awl}
A.~Kitaev and S.~J. Suh, ``{The soft mode in the Sachdev-Ye-Kitaev model and
  its gravity dual},'' {\em JHEP} {\bf 05} (2018) 183,
  \href{http://xxx.lanl.gov/abs/1711.08467}{{\tt 1711.08467}}.

\bibitem{Susskind:1994sm}
L.~Susskind and J.~Uglum, ``{Black hole entropy in canonical quantum gravity
  and superstring theory},'' {\em Phys. Rev.} {\bf D50} (1994) 2700--2711,
  \href{http://xxx.lanl.gov/abs/hep-th/9401070}{{\tt hep-th/9401070}}.

\bibitem{Lewkowycz:2013laa}
A.~Lewkowycz and J.~Maldacena, ``{Exact results for the entanglement entropy
  and the energy radiated by a quark},'' {\em JHEP} {\bf 05} (2014) 025,
  \href{http://xxx.lanl.gov/abs/1312.5682}{{\tt 1312.5682}}.

\bibitem{Srednicki:1993im}
M.~Srednicki, ``{Entropy and area},'' {\em Phys. Rev. Lett.} {\bf 71} (1993)
  666--669, \href{http://xxx.lanl.gov/abs/hep-th/9303048}{{\tt
  hep-th/9303048}}.

\bibitem{Casini:2013rba}
H.~Casini, M.~Huerta, and J.~A. Rosabal, ``{Remarks on entanglement entropy for
  gauge fields},'' {\em Phys. Rev.} {\bf D89} (2014), no.~8 085012,
  \href{http://xxx.lanl.gov/abs/1312.1183}{{\tt 1312.1183}}.

\bibitem{Soni:2015yga}
R.~M. Soni and S.~P. Trivedi, ``{Aspects of Entanglement Entropy for Gauge
  Theories},'' {\em JHEP} {\bf 01} (2016) 136,
  \href{http://xxx.lanl.gov/abs/1510.07455}{{\tt 1510.07455}}.

\end{thebibliography}\endgroup

\end{document}